\documentclass[12pt]{article}
\usepackage{latexsym}
\usepackage{amsmath}
\usepackage{bm}
\usepackage{graphicx,color}
\usepackage{amssymb}
\usepackage{cite}
\usepackage{ulem}

\renewcommand{\theequation}{\thesection.\arabic{equation}}

\makeatletter
\newcommand\xleftrightarrow[2][]{%
  \ext@arrow 9999{\longleftrightarrowfill@}{#1}{#2}}
\newcommand\longleftrightarrowfill@{%
  \arrowfill@\leftarrow\relbar\rightarrow}
\makeatother
\makeatletter
\newcounter{subeqncnt}
\def\thesubeqncnt{\alph{subeqncnt}}
\def\subequations{\begingroup%
\stepcounter{equation}\edef\@tempa{\theequation}%
\let\c@equation\c@subeqncnt\c@subeqncnt\z@
\edef\theequation{\@tempa\noexpand\thesubeqncnt}}

\makeatother
\title{Common Hirota Form B\"{a}cklund Transformation \\
 for the Unified Soliton System }
\author{Masahito Hayashi\thanks{masahito.hayashi@oit.ac.jp}\\
Osaka Institute of Technology, Osaka 535-8585, Japan\\
Kazuyasu Shigemoto\thanks{shigemot@tezukayama-u.ac.jp} \\
Tezukayama University, Nara 631-8501, Japan\\
Takuya Tsukioka\thanks{tsukioka@bukkyo-u.ac.jp}\\
Bukkyo University, Kyoto 603-8301, Japan\\
}
\date{\empty}

\begin{document}

\maketitle
\abstract{
We study to unify soliton systems, KdV/mKdV/sinh-Gordon, through 
SO(2,1) $\cong$ GL(2,$\mathbb R$) $\cong$ M\"{o}bius group point of view, 
which might be a keystone to exactly solve some special 
non-linear differential equations. 

If we construct the $N$-soliton solutions 
through the KdV type B\"{a}cklund transformation, we can transform 
different KdV/mKdV/sinh-Gordon equations and the B\"{a}cklund 
transformations of the standard 
form into the same common Hirota form  and the same common B\"{a}cklund 
transformation except the equation which has the time-derivative term. 
The difference is only the time-dependence and the main structure 
of the $N$-soliton solutions  has the same common form for 
KdV/mKdV/sinh-Gordon systems. Then the $N$-soliton solutions 
for the sinh-Gordon equation is obtained just by the replacement
from KdV/mKdV $N$-soliton solutions.

We also give general addition formulae coming from the KdV type 
B\"{a}cklund transformation which plays not only 
an important role to construct  the 
trigonometric/hyperbolic $N$-soliton solutions  but 
also an essential role to construct the 
elliptic $N$-soliton solutions. 
In contrast to the KdV type B\"{a}cklund transformation,  the  
well-known mKdV/sinh-Gordon type 
B\"{a}cklund transformation gives 
the non-cyclic symmetric $N$-soliton solutions. 
We give an explicit non-cyclic symmetric 3-soliton 
solution for KdV/mKdV/sinh-Gordon equations.
}

%%%%%%%%%%
\section{Introduction} 

\setcounter{equation}{0}

Studies of soliton systems have a long history. 
The discovery of the soliton system by the 
inverse scattering method~\cite{Gardner,Lax,Zakhrov}  has given 
the breakthrough to exactly solve some special non-linear equations. 
There have been many interesting 
developments to understand soliton systems  
such as the AKNS formulation~\cite{Ablowitz,Sasaki}, 
the B\"{a}cklund
transformation~\cite{Wahlquist,Wadati1,Wadati2,Hirota1}, 
the Hirota equation~\cite{Hirota1,Hirota2,Hirota3,Hirota4,Hirota5}, 
the Sato theory~\cite{Sato},
the vertex construction of the soliton solution~\cite{Lepowsky,Date}, 
and the Schwarzian type mKdV/KdV equation~\cite{Weiss}. 
For the construction of $N$-soliton solutions 
of various soliton equations, see the Wawzaz's nice 
 textbook\cite{Wazwaz}.
Even now the soliton theory is quite actively studied in applying 
to the various non-linear phenomena such as (3+1)-dimensional 
lump solution and so on. For example, see Kaur and/or
Wazwaz's recent interesting papers~\cite{Kaur1-1,Kaur1-2,Kaur1-3,Kaur2}. 

In our recent papers,  
we have studied to unify soliton systems such as 
KdV/mKdV/sinh-Gordon equations from 
SO(2,1) $\cong$ GL(2,$\mathbb R$) $\cong$ M\"{o}bius 
group point of view~\cite{Hayashi1,Hayashi2}.
We expect that the various 
approaches above ~\cite{Gardner, Lax,Zakhrov, Ablowitz,Sasaki,
Wahlquist,Wadati1,Wadati2,Hirota1,Hirota2,Hirota3,
Hirota4,Hirota5,Sato,Lepowsky,Date,Weiss} are connected 
through the Lie group. 
We have also formulated soliton systems in a unified manner 
through the Einstein manifold of ${\rm AdS_2}$ in the 
Riemann geometry, which has  SO(2,1) Lie group 
structure~\cite{Hayashi3}. 

We refer a soliton system as that for special types 
of non-linear differential equations, 
which have not only exact solutions but also  
$N$-soliton solutions constructed systematically from $N$ 
pieces of 1-soliton solutions via  
algebraic addition formulae coming 
from the B\"{a}cklund transformation.  
As a result, an expression of
the $N$-soliton solutions becomes a rational function of 
polynomial of many 1-soliton solutions. 
In the representation of the addition formula of 
 SO(2,1) $\cong$ GL(2,$\mathbb R$) $\cong$ M\"{o}bius 
group, algebraic functions such as 
trigonometric/hyperbolic/elliptic 
functions\footnote{In the representation of the 
addition formula of the SO(3) group, the elliptic function comes 
out~\cite{Shigemoto1,Shigemoto2}.} come out.
We consider SO(2,1) $\cong$ GL(2,$\mathbb R$) $\cong$ M\"{o}bius group
 as the keystone for the soliton system.
In the group theoretical point of view, 
we can connect and unify various approaches for soliton systems. 
As the M\"{o}bius group is the rational transformation, 
it is natural to use rational Hirota variables.
Furthermore, as the B\"{a}cklund transformation can be considered
as the self-gauge transformation, it is natural  to use 
B\"{a}cklund transformation as some addition formula of the 
M\"{o}bius group in our Lie group approach.

The B\"{a}cklund transformation goes back 
to Bianchi~\cite{Bianchi}
 for the sine-Gordon equation. It is one of the strong tools to 
construct $N$-soliton solutions. For the old and recent development
of the B\"{a}cklund transformation, see the Rogers-Shadowick's and 
the Rogers-Schief's nice textbooks~\cite{Rogers1,Rogers2}.
The recent hot topics of the B\"{a}cklund transformation is the
application of B\"{a}cklund transformation to the integrable 
defect~\cite{Bowcock,Corrigan,Gomes,Spano}. 

In this paper, 
$N$-soliton solutions would be categorized in terms of two types of 
the B\"{a}cklud transformation. 
We show  one is the well-known KdV type B\"{a}cklud transformation
that provides cyclic symmetric $N$-soliton solutions, while 
another is the well-known mKdV/sinh-Gordon type 
B\"{a}cklund transformation that gives non-cyclic symmetric solutions. 
We also give a general addition formula of the KdV
type B\"{a}cklund transformation.  
An explicit non-cyclic symmetric 3-soliton solution for  
KdV/mKdV/sinh-Gordon equation would be exposed. 
We are interested in the mathematical structure of the integrable 
soliton system, which has $N$-soliton solutions, we did not 
mention the physical applications in this paper.

%%%%%%%%%%
\section{Hirota %type soliton 
forms and their  
B\"{a}cklund transformations} 
\setcounter{equation}{0}

\subsection{KdV equation} 
The KdV equation is given by
\begin{eqnarray}
u_t-u_{xxx}+6 u u_x=0 .
\label{2e1}
\end{eqnarray}
Introducing the $\tau$-function by $u=z_x=-2\left(\log \tau \right)_{xx}$, 
the KdV equation becomes 
\begin{eqnarray}
\frac{\partial}{\partial x}
\left[\frac{(-D_{t}D_{x}+D_x^4)\tau \cdot \tau}{\tau^2}\right]=0 ,
\label{2e2}
\end{eqnarray}
where $D_t, D_x$ are Hirota derivatives defined by 
$D^k_x f(x) \cdot g(x)
   =f(x) (\overleftarrow{\partial_x}-\overrightarrow{\partial_x})^k g(x)$.
Then the KdV equation turns to be so-called Hirota form
\begin{eqnarray}
(-D_{t}D_{x}+D_x^4)\tau \cdot \tau=C_1 \tau^2 ,
\label{2e3}
\end{eqnarray}
with $C_1$ as an integration constant. 
The $C_1 \ne 0$ case 
corresponds to the elliptic soliton case.
\footnote{ 
In the static case, we take the $\tau$-function 
as the Weierstrass's $\sigma$-function, 
then $D_x^4 \tau \cdot \tau=C_1\tau^2$ becomes $\wp_{xx}=6\wp^2-C_1/2$, 
which means that $C_1=g_2$ in the standard notation.
} 
Here we take the
special case i.e.\  $C_1=0$ to consider only the trigonometric/hyperbolic
soliton solution, and we consider the special KdV equation in the form 
\begin{eqnarray}
(-D_{t}D_{x}+D_x^4)\tau \cdot \tau=0  .
\label{2e4}
\end{eqnarray}
One soliton solution for this special Hirota type KdV equation is given by
$$
\tau=1+{\rm e}^{X_i}, \quad {\rm with} \quad X_i=a_i x +a_i^3 t +c_i. 
$$
The Hirota type B\"{a}cklund transformations in this case 
are given by 
\begin{subequations} 
\begin{align}
(-D_t +\frac{3a^2}{4} D_x+D_x^3) \tau' \cdot \tau&=0,
\label{2e5}\\
D_{x}^2 \ \tau' \cdot \tau-\frac{a^2}{4}\tau' \tau&=0.
\label{2e6}
\end{align}
\end{subequations}

\vspace*{-4mm}
\noindent
In fact, using the following relation~\cite{Hirota1}, 
\begin{align}
&\left[(-D_{t}D_{x}+D_x^4)\tau \cdot \tau \right]{\tau'}^2
-{\tau}^2 \left[(-D_{t}D_{x}+D_x^4)\tau' \cdot \tau' \right]
\nonumber\\
&
= -2D_x\left[\left((-D_t +\frac{3a^2}{4} D_x+D_x^3) 
\tau' \cdot \tau \right)\cdot \tau' \tau 
+3\left(D_x \tau' \cdot \tau \right) 
\cdot \left(D_x^2 \tau' \cdot \tau -\frac{a^2}{4}\tau' \tau \right) 
\right],
\label{2e7}
\end{align}
we can show 
that if $\tau$ is the solution of Eq.(\ref{2e4}) and 
if we use Eqs.(\ref{2e5}) and (\ref{2e6}) as the B\"{a}cklund transformations,
then $\tau'$ satisfies 
\begin{eqnarray}
(-D_{t}D_{x}+D_x^4)\tau' \cdot \tau'=0,
\label{2e8}
\end{eqnarray}
which means that $\tau'$ is a new solution.

We now show that 
the Hirota type B\"{a}cklund transformation  
Eq.(\ref{2e6}) relates to the following
well-known KdV type B\"{a}cklund transformation 
\begin{eqnarray}
z_x+z'_x=-\frac{a^2}{2}+\frac{1}{2}(z-z')^2.
\label{2e10}
\end{eqnarray}
Writing down Eq.(\ref{2e6}) more explicitly, 
\begin{equation}
D_x^2 \tau' \cdot \tau
=\tau' \tau_{xx} -2 \tau'_x \tau_x+\tau'_{xx} \tau
=\frac{a^2}{4}\tau' \tau, 
\label{2e11}
\end{equation}
and defining $z=-2\dfrac{\tau_x}{\tau}$ and 
$z'=-2\dfrac{\tau'_x}{\tau'}$, 
we can organize Eq.(\ref{2e10}) as 
\begin{align}
&
z'_x+z_x+\frac{a^2}{2}-\frac{1}{2}(z'-z)^2 
=-2 \frac{\tau'_{xx}}{\tau'}+2 \frac{\tau'^2_{x}}{\tau'^2}
-2 \frac{\tau_{xx}}{\tau}
+2 \frac{\tau^2_{x}}{\tau^2} +\frac{a^2}{2}
-2\left(\frac{\tau'_x}{\tau'}-\frac{\tau_x}{\tau}\right)^2
\nonumber\\
&=-\frac{2}{\tau' \tau}\left(  \tau' \tau_{xx} -2 \tau'_x \tau_x+\tau'_{xx} \tau
-\frac{a^2}{4}\tau' \tau \right)
\nonumber 
\\
&=-\frac{2}{\tau' \tau}
\left( D_x^2 \tau' \cdot \tau-\frac{a^2}{4}.\tau' \tau \right),
\label{2e12}
\end{align}
which leads the following equivalence 
\begin{equation}
D_x^2 \tau' \cdot \tau=\frac{a^2}{4} \tau' \tau \quad 
\Longleftrightarrow \quad 
z'_x+z_x=-\frac{a^2}{2}+\frac{1}{2}(z'-z)^2  .
\label{2e13}
\end{equation}

In the previous paper~\cite{Hayashi1}, we make the connection 
between the KdV equation  and the mKdV equation 
through the Miura transformation $u=\pm v_x+v^2$ 
with the common Hirota type variables $f$ and $g$, that is,
$u=-2(\log \tau )_{xx}$, $\tau=f\pm g$ in the KdV equation 
and $v=w_x$, $\tanh w/2=g/f$ in the mKdV equation. 
In order to connect the KdV equation with the mKdV 
equation, 
we would like to take variables $f$ and $g$ as 
$\tau=f \pm g$, $\tau'=f' \pm g'.$ 
For the $N$-soliton solution, $f$ and $g$ are an even and an  
odd part of a $N$-soliton solution 
under changing an overall sign of each 1-soliton solution. 
We refer $f$ and $g$ as Hirota form variables.  
In order to construct $N$-soliton solutions, only one of the
B\"{a}cklund transformations 
Eq.(\ref{2e6}) is enough, which is given by
\begin{eqnarray}
D_{x}^2 \ (f' \pm g') \cdot (f \pm g)=\frac{a^2}{4} (f' \pm g')(f \pm g)   .
\label{2e9}
\end{eqnarray}

We can simplify Eq.(\ref{2e4}) by using $f$ and $g$ variables.
By using the soliton number unchanging self B\"{a}cklund transformation,
i.e.\ $f'=f$, $g'=-g$, and $a=0$ in Eq.(\ref{2e9}), we have
\begin{eqnarray}
D_x^2 (f \cdot f -g \cdot g)=0.
\label{2add1}
\end{eqnarray}
While by using $p=f+g$ and $q=f-g$, we obtain an identity 
\begin{align}
&\left( (-D_t D_x+D_x^4)p \cdot p\right)q^2
-p^2 \left( (-D_t D_x+D_x^4) q \cdot q \right) \nonumber\\
&= D_x\left[ 2\left( (-D_t+D_x^3) p \cdot q \right) \cdot p q 
+12 \left(D_x^2(f \cdot f-g\cdot g)\right) \cdot \left(D_x(f\cdot g) \right) \right] .
\label{2add2}
\end{align}
Since we have $ (-D_t D_x+D_x^4)p \cdot p=0$ and 
$ (-D_t D_x+D_x^4)q \cdot q=0$ from Eq.(\ref{2e4}) 
with $\tau=f\pm g$, if we use 
Eq.(\ref{2add1}), we have
$(-D_t+D_x^3) p \cdot q=-2(-D_t+D_x^3)(f \cdot g)=0$.
In this way, Eq.(\ref{2e4}) is simplified in the following forms
\begin{subequations} 
\begin{align}
(-D_t+D_x^3)f \cdot g&=0, 
\label{2add3}\\
D_x^2 (f \cdot f -g \cdot g)&=0.
\label{2add4}
\end{align}
\end{subequations}%
We call Eq.(\ref{2add4}) as a structure equation, which determines the structure 
of $N$-soliton solutions.
While we refer Eq.(\ref{2add3}) as a dynamical equation, which yields 
time dependence of $N$-soliton solutions. 
In next subsection, we will see that these equations are the same 
as those in the special mKdV equation.

%%%%%%%%%%%%%%%%%
\subsection{mKdV equation} 

The mKdV equation is given by
\begin{eqnarray}
v_t-v_{xxx}+6 v^2 v_x=0.
\label{2e14}
\end{eqnarray}
Defining $v=w_x$ and $\tanh(w/2)=g/f$, 
we get 
\begin{equation}
\frac{ (-D_{t}+D_x^3)f \cdot g}{D_x f \cdot g}
=3\frac{ D_x^2(f \cdot f-g \cdot g)}{f^2-g^2}   .
\label{2e15}
\end{equation}

We now consider the following special case 
\begin{subequations} 
\begin{align}
(-D_{t}+D_x^3)f \cdot g&=0, 
\label{2e16}\\
D_x^2(f \cdot f-g \cdot g)&=0.
\label{2e17}
\end{align}
\end{subequations}

\vspace*{-5mm}
\noindent
Then we have the common structure equation Eq.(\ref{2e17}) 
in the mKdV equation as that of Eq.(\ref{2add4}) in the KdV equation.
Further we have the common dynamical equation Eq.(\ref{2e16}) in 
the mKdV equation as that of Eq.(\ref{2add3}) in the KdV equation. 
 
One soliton solution for this special Hirota type mKdV equation
(\ref{2e16}) and (\ref{2e17}) is given by
$$
f=1, \qquad g={\rm e}^{X_i}, \quad {\rm with}\quad X_i=a_i x+a_i^3 t+c_i.
$$

The B\"{a}cklund transformation for 
the structure equation (\ref{2e17}) is given by~\cite{Hirota1}
\begin{subequations}
\begin{align}
D_x(f'-g')\cdot(f+g)&=-\frac{a}{2} (f'+g')(f-g)  ,
\label{2e18}\\
D_x(f'+g')\cdot(f-g)&=-\frac{a}{2} (f'-g')(f+g) ,
\label{2e19}
\end{align}
\end{subequations}

\vspace*{-4mm}
\noindent
by using the following relations. 
Taking Eqs.(\ref{2e18}) and (\ref{2e19}) into account, we have 
a relation 
\begin{align}
&
\left[D_x^2(f'+g')\cdot(f'-g')\right](f+g)(f-g)
-(f'+g')(f'-g')\left[D_x^2(f+g)\cdot(f-g)\right]
\nonumber\\
&
=
D_x\left[\left( D_x(f'+g') \cdot (f-g)+\frac{a}{2}(f'-g')(f+g)\right) \cdot (f'-g')(f+g)
\right. 
\nonumber\\
&
\quad\left.
-\left( D_x(f'-g') \cdot (f+g)+\frac{a}{2}(f'+g')(f-g)\right) \cdot (f'+g')(f-g) \right] 
\nonumber\\
&
\quad
+D_x\left[-\frac{a}{2}(f'-g')(f+g)\cdot (f'-g')(f+g)
+\frac{a}{2}(f'+g')(f-g)\cdot (f'+g')(f-g)\right] 
\nonumber\\
&= 
D_x\left[\left( D_x(f'+g') \cdot (f-g)+\frac{a}{2}(f'-g')(f+g)\right) \cdot (f'-g')(f+g)
\right. \nonumber\\
&
\quad 
\left.
-\left( D_x(f'-g') \cdot (f+g)+\frac{a}{2}(f'+g')(f-g)\right) 
\cdot (f'+g')(f-g) \right], 
\label{2e20}
\end{align}
where we have used $D_x F \cdot F=F_x F-F F_x = 0$. 
This relation means that 
if  Eqs.(\ref{2e17}), (\ref{2e18}), and (\ref{2e19}) are satisfied, we have
$D_x^2(f '\cdot f'-g '\cdot g')=0$, that is, 
if the set $(f,g)$ is a solution, 
the set $(f',g')$ produces a new solution 
by using the B\"{a}cklund transformation. 
  
We can find equivalent forms for 
the B\"acklund transformations (\ref{2e18}) and
(\ref{2e19})~\cite{Hirota1}.  
First, we consider the following relation
\begin{align}
&\frac{D_x^2(f'+g') \cdot (f+g)}{(f'+g') (f+g)}
-\frac{D_x^2(f'-g') \cdot (f-g)}{(f-+g') (f-g)} 
\nonumber\\
&= \frac{1}{(f'^2-g'^2)(f^2-g^2)}
D_x\left[ \left(D_x (f' +g') \cdot (f-g)+\frac{a}{2}(f'-g')(f+g)\right)
\cdot (f'-g')(f+g)\right. 
\nonumber\\
&\hspace*{45mm}
+\left. (f'+g')(f-g) \cdot 
\left(D_x (f' -g') \cdot (f+g)+\frac{a}{2}(f'+g')(f-g)\right) \right]
\nonumber 
\\
&=0, 
\label{2e22}
\end{align}
where we have used the B\"acklund transformations 
(\ref{2e18}) and (\ref{2e19}). 
Secondly, we obtain 
\begin{align}
&\frac{D_x^2(f'+g') \cdot (f+g)}{(f'+g') (f+g)}
+\frac{D_x^2(f'-g') \cdot (f-g)}{(f-g') (f-g)}-\frac{a^2}{2}
\nonumber 
\\
&=\left[ \frac{D_x^2(f' \cdot f'-g' \cdot g')}{(f'^2-g'^2) }
+\frac{D_x^2(f \cdot f-g \cdot g)}{(f^2-g^2)}     \right]
\nonumber\\
&\quad 
+2\left[ \left(\frac{D_x(f'+g') \cdot (f-g)}{(f'-g') (f+g)}\right) 
\left( \frac{D_x(f'-g') \cdot (f+g)}{(f'+g') (f-g)} \right)-\frac{a^2}{4}
\right]
\nonumber 
\\
&=
2\left[ \left(\frac{D_x(f'+g') \cdot (f-g)}{(f'-g') (f+g)}\right) 
\left( \frac{D_x(f'-g') \cdot (f+g)}{(f'+g') (f-g)} \right)-\frac{a^2}{4}
\right]
\nonumber 
\\
&=0 ,
\label{2e23}
\end{align}
where we have used  
the structure equations 
$D_x^2(f'\cdot f'-g' \cdot g')=0$ and $D_x^2(f\cdot f-g \cdot g)  =0$
and also the B\"acklund transformations (\ref{2e18}) and (\ref{2e19}). 
Combining Eqs.(\ref{2e22}) and (\ref{2e23}), we 
arrive at 
\begin{equation}
D_{x}^2 (f'\pm g')\cdot (f \pm g)=\frac{a^2}{4}(f'\pm g')(f \pm g).  
\label{2e21}
\end{equation}
Then we have the common Hirota form B\"{a}cklund 
transformation  Eq.(\ref{2e21}) in the mKdV equation 
as that of Eq.(\ref{2e9}) in the KdV equation. 
This is the reason why we call this as the common KdV type 
Hirota form B\"{a}cklund transformation. 
 
Conversely, if  Eq.(\ref{2e21}) is satisfied, we have
\begin{align}
&D_x\left[ \left(D_x (f' +g') \cdot (f-g)
+\frac{a}{2}(f'-g')(f+g)\right)\cdot (f'-g')(f+g)\right.
\nonumber\\
&\qquad 
\left. + (f'+g')(f-g) \cdot \left(D_x (f' -g') \cdot (f+g)
+\frac{a}{2}(f'+g')(f-g)\right) \right]=0  , 
\label{2e25}\\
& \left(\frac{D_x(f'+g') \cdot (f-g)}{(f'-g') (f+g)}\right) 
\left( \frac{D_x(f'-g') \cdot (f+g)}{(f'+g') (f-g)} \right)-\frac{a^2}{4}=0,
 \label{2e26}
\end{align}
which give Eq.(\ref{2e18}) and Eq.(\ref{2e19}) 
by properly choosing the sign of $a$.
Then we conclude the equivalence   
\begin{equation}
\frac{D_x(f' \pm g') \cdot (f \mp g)}{(f' \mp g') (f \pm g)}=-\frac{a}{2} 
\quad\Longleftrightarrow\quad 
\frac{D_x^2(f'\pm g') \cdot (f \pm g)}{(f' \pm g') (f \pm g)}=\frac{a^2}{4}  .
\label{2e27}
\end{equation}
The Eq.(\ref{2e21}) is the Hirota type B\"{a}cklund transformation for the
special mKdV structure equation Eq.(\ref{2e17}). 
 
Now we focus on yet another 
mKdV type B\"{a}cklund transformation~\cite{Hirota1} 
\begin{equation}
 w'_x+w_x=a \sinh (w'-w). 
\label{2e28}
\end{equation}
From Eqs.(\ref{2e18}) and (\ref{2e19}), we can obtain (\ref{2e28}), 
while the opposite is not always true:  
\begin{equation}
D_x (f' \pm g') \cdot (f \mp g)=-\frac{a}{2}(f' \mp g')(f \pm g)
 \quad\Longrightarrow\quad w'_x+w_x=a \sinh (w'-w). 
\label{2e32}
\end{equation}
We can show the relation above in the following manner. 
Using
$$
\tanh\frac{w}{2}=\frac{g}{f}, \quad \
\sinh w=\frac{2f g}{f^2-g^2}, \quad \cosh w=\frac{f^2+g^2}{f^2-g^2},
$$ 
and their counterparts for $(w', f', g')$, 
we have
$$
w_x=\frac{2(f g_x-f_x g)}{f^2-g^2}
=-\frac{D_x(f-g)\cdot (f+g)}{f^2-g^2}, 
$$
and those for $(w', f', g')$. 
Then we have a relation
\begin{align}
&\left[ D_x(f'+g') \cdot (f-g)+\frac{a}{2} (f'-g')(f+g) \right]
 (f'-g')(f+g)
\nonumber\\
&\quad 
-(f'+g')(f-g)\left[ D_x(f'-g') \cdot (f+g)+\frac{a}{2} (f'+g')(f-g)  \right] 
\nonumber\\
&= \left[ D_x(f'+g') \cdot (f'-g') \right] (f-g)(f+g)
- (f'+g')(f'-g')\left[ D_x(f-g) \cdot (f-g) \right] 
\nonumber\\
&\quad 
+\frac{a}{2} \left[ (f'-g')^2(f+g)^2- (f'+g')^2(f-g)^2\right]
\nonumber\\
&=(f'^2-g'^2)(f^2-g^2)\left[ w'_x+w_x
+a\left(\frac{f'^2+g'^2}{f'^2-g'^2}\frac{2f g}{f^2-g^2}
-\frac{2 f' g'}{f'^2-g'^2}\frac{f^2+g^2}{f^2-g^2}\right)\right]
\nonumber\\
&= (f'^2-g'^2)(f^2-g^2)\left[ w'_x+w_x+a(\cosh(w') \sinh(w)-\sinh(w') \cosh(w) \right]
\nonumber\\
&= (f'^2-g'^2)(f^2-g^2)\left[ w'_x+w_x-a \sinh(w'-w) \right]  ,
\label{2e31}
\end{align}
which means we have Eq.(\ref{2e28}) from Eqs.(\ref{2e18}) and 
(\ref{2e19}), but the opposite is not always shown. 
In fact, Eq.(\ref{2e28}) is  
the B\"{a}cklund transformation of the original mKdV equation Eq.(\ref{2e15}) 
but not the B\"{a}cklund transformation 
of the special  mKdV equations Eqs.(\ref{2e16})  
and  (\ref{2e17}). 

By the KdV type Hirota form B\"{a}cklund 
transformation Eq.(\ref{2e21}), 
we have the cyclic symmetric $N$-soliton solutions. 
On the other hand, by the mKdV type B\"{a}cklund transformation
Eq.(\ref{2e28}), we have the non-cyclic symmetric $N$-soliton solutions. 
In section 4, 
we give an explicit non-cyclic symmetric 3-soliton solution from 
mKdV type B\"{a}cklund transformation Eq.(\ref{2e28}).

%%%%%%%%%%%%%%%%%%%%%%%%%%
\subsection{sinh-Gordon equation} 
The sinh-Gordon equation is given by
\begin{eqnarray}
\theta_{xt}=\sinh \theta  .
\label{2e33}
\end{eqnarray}
Defining $\tanh(\theta/4)=g/f$, we obtain 
\begin{equation}
\frac{ D_{t}D_x\ f \cdot g}{f g}-1
=\frac{D_{t} D_{x}(f \cdot f+g \cdot g)}{f^2+g^2}  .
\label{2e34}
\end{equation}
We here consider the special case: 
\begin{subequations} 
\begin{align}
&D_{t}D_x\ f \cdot g=f g , 
\label{2e35}\\
&D_{t} D_{x}(f \cdot f+g \cdot g)=0  .
\label{2e36}
\end{align}
\end{subequations}

\vspace*{-4mm}
\noindent
Taking the following relation into account, 
\begin{align}
&D_x\left[ D_t D_x(f \cdot f +g \cdot g) \cdot (f^2+g^2)
-4\left(D_t D_x(f \cdot g) -f  g\right) \cdot fg \right]  
\nonumber\\
&=D_t\left[ \left( D_x^2(f \cdot f - g \cdot g)\right) \cdot
 (f^2-g^2)\right], 
\label{2e37}
\end{align}
we take
\begin{subequations}
\begin{eqnarray}
&&D_{t}D_x\ f \cdot g=f g  , 
\label{2e38}\\
&&D_{x}^2(f \cdot f-g \cdot g)=0  , 
\label{2e39}
\end{eqnarray}
\end{subequations}

\vspace*{-4mm}
\noindent
as the special sinh-Gordon equation instead of Eqs.(\ref{2e35}) 
and (\ref{2e36}). 
The above structure equation Eq.(\ref{2e39}) in 
the sinh-Gordon equation is the same as that 
of Eq.(\ref{2add4}) in the KdV equation and Eq.(\ref{2e17}) in the 
mKdV equation. 
Then, applying the same method as 
that of the mKdV equation,  
we have the common KdV type Hirota form 
B\"{a}cklund transformation Eqs.(\ref{2e18}) and (\ref{2e19}) , 
and equivalently Eq.(\ref{2e21}) for KdV/mKdV/sinh-Gordon equations. 

One soliton solution for this special type sinh-Gordon equation 
is given by
$$
f=1, \qquad 
g={\rm e}^{\hat{X}_i}, \quad {\rm with}\quad 
\hat{X}_i=a_i x+t/a_i +c_i. 
$$

From Eqs.(\ref{2e18}) and (\ref{2e19}), 
we have another mKdV type B\"{a}cklund transformation 
by replacing $w \rightarrow \theta/2$ in the mKdV
type B\"{a}cklund transformation Eq.(\ref{2e28}). This is because the relation
$\tanh(w/2)=g/f$ in the mKdV equation corresponds to
$\tanh(\theta/4)=g/f$ in the sinh-Gordon equation. 
Then from Eqs.(\ref{2e18}) and (\ref{2e19}), we have
\begin{equation}
\frac{\theta'_x}{2} +\frac{\theta_x}{2}
=a\sinh \left(\frac{\theta'}{2}-\frac{\theta}{2}\right)  ,
\label{2e43}
\end{equation}
but the opposite is not always satisfied. 
In fact, Eq.(\ref{2e43}) is the 
B\"{a}cklund transformation for the original sinh-Gordon equation Eq.(\ref{2e34}) 
but not the B\"{a}cklund transformation of the special sinh-Gordon equation 
Eqs.(\ref{2e38}) and (\ref{2e39}).

%%%%%%%%%%%%%%%%%%%%%%%%
\subsection{
Cyclic symmetric $N$-soliton solutions 
via Hirota form \\
B\"{a}cklund
transformations}

Let us first summarize our findings in the previous subsections.  
By using the Hirota form variables $f$ and $g$,  we can treat 
the special KdV/mKdV/sinh-Gordon equations in a unified manner:  
\begin{align}
\text{a)} &\quad \mbox{KdV\ Eq.}: \ 
u=z_x=-2\left(\log \tau \right)_{xx},\quad \tau=f \pm g ,  
\label{2e44}\\
\text{b)} &\quad \mbox{mKdV\ Eq.}:\ 
v=w_x,\quad \tanh \frac{w}{2}=\frac{g}{f} , 
\label{2e45}\\
\text{c)} &\quad \mbox{sinh-Gordon\ Eq.}:\ 
\tanh \frac{\theta}{4} =\frac{g}{f}.  
\label{2e46}
\end{align}
The well-known KdV type B\"{a}cklund transformation is equivalent
to the KdV type Hirota form B\"{a}cklund transformation: 
\begin{align}
&z'_x+z_x=-\frac{a^2}{2}+\frac{(z'-z)^2}{2} 
\nonumber 
\\
&\Longleftrightarrow\quad 
D_x^2 (f' \pm g') \cdot (f \pm g)=\frac{a^2}{4} (f' \pm g') 
\cdot (f \pm g). 
\label{2e47}
\end{align}
We have the common KdV type 
Hirota form B\"{a}cklund transformation Eq.(\ref{2e47})
for the special KdV equation Eq.(\ref{2add3}) and Eq.(\ref{2add4}), 
for the special mKdV equation  Eqs.(\ref{2e16}) and (\ref{2e17}), and for 
the special sinh-Gordon equation Eqs.(\ref{2e38}) and (\ref{2e39}) 
for the common structure equation Eqs.(\ref{2add4}), 
(\ref{2e17})  and (\ref{2e39}). 
Another mKdV type B\"{a}cklund
transformation Eq.(\ref{2e28}) is the B\"{a}cklund
transformation of the original mKdV equation Eq.(\ref{2e15}) 
but not the  B\"{a}cklund transformation of the special mKdV equation
 Eq.(\ref{2e16}) and Eq.(\ref{2e17}).

In our previous paper~\cite{Hayashi1}, we have demonstrated how to construct 
$N$-soliton solutions from $N$ pieces of 1-soliton solutions by using 
KdV type B\"{a}cklund transformation Eq.(\ref{2e10}). 
Here we demonstrate how to construct the cyclic $N$-soliton solutions for $N$=2 case.
We start from the addition formula of the B\"{a}cklund transformation,  
\begin{eqnarray}
z_{12}=\frac{a_1^2-a_2^2}{z_1-z_2} ,
\label{2e48}
\end{eqnarray}
where we choose 
$$
z_0=0, \qquad 
z_i=-a_i \tanh X_i/2, %=a_i (1-e^{X_i})/(1+e^{X_i}), 
\quad\mbox{with}\quad X_i=a_i x +a_i^3 t + c_i.
$$
In order to find a KdV two-soliton solution, 
we simply take the space derivative
by using $u_{12}=z_{12, x}$. 
While,  if we want to find a 2-soliton solution 
for the mKdV/sinh-Gordon equation, 
we must know $f_{12}$ and $g_{12}$ from $z_{12}$.
We can find $f_{12}$ and $g_{12}$ from 
$z_{12}=-2 \tau_{12, x}/\tau_{12} +\text{const.}$
 with $\tau_{12}=f_{12} \pm g_{12}$~\cite{Hayashi1}, 
but it becomes complicated for 
the general $N$-soliton solutions. 
However, it is easier to find the $\tau_{12}$-function 
directly from the Hirota equation $(-D_t D_x+D_x^4) \tau_{12} \cdot \tau_{12}=0$ 
in the standard way~\cite{Hirota5,Wazwaz}, which gives 
\begin{equation}
\tau_{12}=f_{12} \pm g_{12}
\label{2e49}
\end{equation}
with
\begin{subequations}
\begin{align}
f_{12}&= 1+\frac{(a_1-a_2)^2}{(a_1+a_2)^2}\, {\rm e}^{X_1}{\rm e}^{X_2} ,
\label{2e50}\\ 
g_{12}&={\rm e}^{X_1}+{\rm e}^{X_2} ,
\label{2e51}
\end{align} 
\end{subequations}

\vspace*{-4mm}
\noindent
where $f$ and $g$ are even and odd parts of the $\tau_{12}$ function under 
${\rm e}^{X_i} \rightarrow -{\rm e}^{X_i}$. 
For a 2-soliton solution of mKdV equation, 
we have $\tanh (w_{12}/2)=g_{12}/f_{12}$~\cite{Wazwaz}.
For a soliton solution of sinh-Gordon equation, 
using the dynamical equation Eq.(\ref{2e38}),  
we replace
$X_i \rightarrow \hat{X}_i$ with $ \hat{X}_i=a_i x+t/a_i +c_i$, 
because $f=1$, $g={\rm e}^{\hat{X}_i}$ is 
a 1-soliton solution of $D_t D_x f \cdot g=fg$.
Then the 2-soliton solution of sinh-Gordon equation is given by
$\tanh \theta_{12}/4=\hat{g}/\hat{f}$~\cite{Wazwaz}, where 
$\hat{f}_{12}= 1+(a_1-a_2)^2/(a_1+a_2)^2 \, 
{\rm e}^{\hat{X}_1}{\rm e}^{\hat{X}_2}$,
$\hat{g}_{12}= {\rm e}^{\hat{X}_1}+{\rm e}^{\hat{X}_2}$. 

In general, we have the cyclic symmetric $N$-soliton 
solutions~\cite{Wazwaz} by using 
the common KdV type B\"{a}cklund transformation.
 
%%%%%%%%%%
\section{Addition formulae for the common KdV 
type \\
B\"{a}cklund transformation} 
\setcounter{equation}{0}

In our approach, we construct cyclic symmetric $N$-soliton
solutions by an algebraic addition formula coming from the well-known KdV
type B\"{a}cklund transformation, which is equivalent to the  
common KdV type B\"{a}cklund transformation. 
This addition formula is applicable also to construct the 
elliptic $N$-soliton solutions and there will be no other way 
to construct $N$-soliton solutions for the elliptic case~\cite{Hayashi2}.
In order to construct $N$-soliton solutions for trigonometric/hyperbolic/elliptic 
soliton solutions, we give the result of the general addition formula here.

Let us first review to find a 2-soliton solution 
by the common KdV type B\"{a}cklund transformation.
Assuming the commutativity, $z_{12}=z_{21}$, we have
\begin{subequations} 
\begin{align}
 z_{1, x}+z_{0, x}&=-\dfrac{a^2_1}{2}+\dfrac{(z_1-z_0)^2}{2},
\label{3e1}\\
z_{2, x}+z_{0, x}&=-\dfrac{a^2_2}{2}+\dfrac{(z_2-z_0)^2}{2},
\label{3e2}\\
z_{12, x}+z_{1, x}&=-\dfrac{a^2_2}{2}+\dfrac{(z_{12}-z_1)^2}{2},
\label{3e3}\\
z_{12, x}+z_{2, x}&=-\dfrac{a^2_1}{2}+\dfrac{(z_{12}-z_2)^2}{2}.
\label{3e4}
\end{align}  
\end{subequations}

\vspace*{-4mm}
\noindent
Making Eq.(\ref{3e1})$-$Eq.(\ref{3e2})$-$Eq.(\ref{3e3})$+$Eq.(\ref{3e4}),
derivative terms are canceled out and we have
\begin{equation}
z_{12}=z_0+\frac{a^2_1-a^2_2}{z_1-z_2}.
\label{3e5} 
\end{equation}
We can check that Eq.(\ref{3e5}) satisfies 
Eqs.(\ref{3e1})-(\ref{3e4}), which means that it is commutative in this level. 
Recursively, we have 
\begin{equation}
z_{1 2 \cdots,n-2,n-1,n}
=z_{1 2 \cdots,n-2}
+\frac{{a_{n-1}}^2-{a_{n}}^2}{z_{1 2 \cdots,n-2,n-1}-z_{1 2 \cdots,n-2,n}}.
\label{3e6}
\end{equation}

We list various $N$-soliton solutions obtained through the addition formulae: 

%%%%%%%%%%%%
%\subsection{(2+1)-soliton solution}
%The result of the addition formula for (2+1)-soliton solution is given by 
% 
\begin{itemize}
 \item  {\bf (2+1)-soliton solution}
\\
\begin{equation}
z_{12}=z_0+\frac{{a_1}^2-{a_2}^2}{z_{1}-z_{2}}=z_0+\frac{G_{12}}{F_{12}} ,
\label{3e7}
\end{equation}
with
\begin{subequations} 
\begin{align}
F_{12}&=z_1-z_2, 
\label{3e8}\\
G_{12}&=a_1^2-a_2^2 .
\label{3e9}
\end{align}
\end{subequations}
%
%
%%%%%%%%%
%\subsection{3-soliton solution}

%The result of the addition formula for 3-soliton solution is given by 
%
\item {\bf 3-soliton solution}
\\
\begin{equation}
z_{123}=z_1+\frac{{a_2}^2-{a_3}^2}{z_{12}-z_{13}}=\frac{G_{123}}{F_{123}} ,
\label{3e10}
\end{equation}
with
\begin{subequations}
\begin{align}
F_{123}&=({a_1}^2-{a_2}^2)z_3+({a_2}^2-{a_3}^2)z_1
+{(a_3}^2-{a_1}^2)z_2 
\nonumber\\
&=\frac{1}{2!}\sum_{i,j,k=1}^{3}\epsilon^{ijk}(a_i^2-a_j^2)z_k ,
\label{3e11}\\
G_{123}&=-\left(({a_1}^2-{a_2}^2)z_1 z_2
+({a_2}^2-{a_3}^2)z_2 z_3+({a_3}^2-{a_1}^2)z_3 z_1\right) \nonumber\\
&=-\frac{1}{2!}\sum_{i,j,k=1}^{3}\epsilon^{ijk}
(a_i^2-a_j^2)z_i z_j. 
\label{3e12}
\end{align}
\end{subequations}
%
%where $\epsilon^{i_1i_2\cdots i_n}$ is a Levi-Civita symbol with 
%$\epsilon^{12\cdots n}=1$.
%
%%%%%%%%%%
%\subsection{(4+1)-soliton solution} 

%The result of the addition formula for (4+1)-soliton solution is given by 
%
\item {\bf (4+1)-soliton solution}
\\
\begin{equation}
z_{1234}
=z_{12}+\frac{{a_3}^2-{a_4}^2}{z_{123}-z_{124}}
=z_0+\frac{G_{1234}}{F_{1234}} , 
\label{3e13}
\end{equation}
with
\begin{subequations}
\begin{align}
F_{1234}&=\frac{1}{(2!)^2} \sum^4_{i,j,k,l=1} \epsilon^{ijkl}
 ({a_i}^2-{a_j}^2)({a_k}^2-{a_l}^2) z_i z_j , 
\label{3e14}\\
G_{1234}&=-\frac{1}{2!} \sum^4_{i,j,k,l=1}  \epsilon^{ijkl}
 {a_i}^2 {a_j}^2({a_i}^2-{a_j}^2) z_k.  
\label{3e15}
\end{align}
\end{subequations}
%
%
%%%%%%%%%%%%%%%%%%%
%\subsection{5-soliton solution} 
%
%The result of the addition formula for 5-soliton solution is given by 
%
\item {\bf 5-soliton solution}
\\
\begin{equation}
z_{12345}
=z_{123}+\frac{{a_4}^2-{a_5}^2}{z_{1234}-z_{1235}}
=\frac{G_{12345}}{F_{12345}}  ,
\label{3e16}\\
\end{equation}
with
\begin{subequations}
\begin{align}
F_{12345} 
&=\frac{1}{3! 2!}\sum^5_{i,j,k,l,m=1} \epsilon^{ijklm}
 ({a_i}^2-{a_j}^2) \left[({a_k}^2-{a_l}^2)
({a_l}^2-{a_m}^2)({a_m}^2-{a_k}^2)\right] z_i z_j,
\label{3e17}\\
G_{12345}
&=\frac{1}{3! 2!} \sum^5_{i,j,k,l,m=1} \epsilon^{ijklm}
 ({a_i}^2-{a_j}^2)\left[({a_k}^2-{a_l}^2)
({a_l}^2-{a_m}^2)({a_m}^2-{a_k}^2)\right] z_k z_l z_m,  
\label{3e18}
\end{align}
\end{subequations}
where $\epsilon^{i_1i_2\cdots i_n}$ is a Levi-Civita symbol with 
$\epsilon^{12\cdots n}=1$.
\end{itemize}

%%%%%%%%%%%%%%%%%%%
\subsection{General formula}

We first define the following quantity 
\begin{eqnarray}
&&\Lambda(i_1,i_2, \cdots, i_n)
=\sum_{\substack{p,q=1 \\ p<q}}^n(a_{i_p}^2-a_{i_q}^2)  , 
\qquad
\Pi(i_1,i_2, \cdots, i_n)=\prod_{p=1}^{n} a_{i_p}^2  , 
\nonumber
\end{eqnarray}
where we set $\Lambda(i_1,i_2)=1$. 
The general formula is expected to be given in the following form: 
\begin{itemize}
 \item {\bf ((2n)+1)-solution}
\\
\begin{equation}
%\hspace*{-50mm}
%\bullet \quad \text{\underline{((2n)+1)-solution}}:\qquad\qquad 
z_{12\cdots 2n}=z_0+\frac{G_{12\cdots 2n}}{F_{12\cdots 2n}}, 
\label{3e19}
\end{equation}
with
\begin{subequations}
\begin{align}
F_{12\cdots 2n}
&=\frac{1}{(n!)^2} \sum 
\epsilon^{i_1 i_2 \cdots i_n j_1 j_2 \cdots j_n}
\Lambda(i_1,i_2,\cdots, i_n) \Lambda(j_1,j_2,\cdots, j_n) 
z_{i_1} z_{i_2}\cdots z_{i_n}, 
\\  
G_{12\cdots 2n}
&=-\frac{(-1)^n}{n!(n-1)!} \sum 
\epsilon^{i_1 i_2 \cdots i_n j_1 j_2 \cdots j_n}
\Lambda(i_1,i_2,\cdots, i_n) \Pi(i_1,i_2,\cdots, i_n) 
z_{j_1} z_{j_2}\cdots z_{j_{n-1}}.
\end{align}
\end{subequations}%
\item {\bf (2n+1)-solution}
\\
\begin{equation}
%\hspace*{-50mm}\bullet\quad 
%\text{\underline{(2n+1)-solution} }:\qquad\qquad 
z_{12\cdots 2n+1}=\frac{G_{12\cdots 2n+1}}{F_{12\cdots 2n+1}}  ,
\label{3e20}
\end{equation}
with
\begin{subequations}
\begin{align}
F_{12\cdots 2n+1}
&=\frac{1}{n!(n+1)!}
\nonumber 
\\
&\quad\times \sum 
\epsilon^{i_1 i_2 \cdots i_n j_1 j_2 \cdots j_n j_{n+1}}
\Lambda(i_1,i_2,\cdots, i_n) \Lambda(j_1,j_2,\cdots, j_n,j_{n+1}) 
z_{i_1} z_{i_2}\cdots z_{i_n}  , 
\\
G_{12\cdots 2n+1}
&=\frac{(-1)^n}{n!(n+1)!}
\nonumber 
\\
&\quad\times \sum 
\epsilon^{i_1 i_2 \cdots i_n j_1 j_2 \cdots j_n j_{n+1}}
\Lambda(i_1,i_2,\cdots, i_n) \Lambda(j_1,j_2,\cdots, j_n,j_{n+1})
z_{j_1} z_{j_2}\cdots z_{j_{n+1}} .   
\end{align}
\end{subequations}
\end{itemize}

\vspace*{-4mm}
\noindent
We have checked these formulae up to $z_{1234567}$ by Mathematica. 

%%%%%%%%%%%
\section{Non-cyclic symmetric 3-soliton solutions 
of the mKdV equation} 
\setcounter{equation}{0}

Here we consider that another mKdV type B\"{a}cklund 
transformation Eq.(\ref{2e28}) of the original mKdV equation gives 
non-cyclic symmetric soliton solutions. 
We demonstrate on that by constructing a 3-soliton solution.
  
Another mKdV type B\"{a}cklund transformation of the mKdV 
equation is given by\cite{Wadati1, Wadati2} 
\begin{align}
w_x'+w_x&=a \sinh(w'-w),
\label{4e1}\\
w_t'+w_t&=-2a^2 w_x-2a w_{xx} \cosh(w'-w)
+(a^3-2a {w_x}^2) \sinh(w'-w).
\label{4e2}\     
\end{align}
This B\"{a}cklund transformation can be considered as 
a self gauge transformation of the 
GL(2,$\mathbb R$) in the AKNS formalism~\cite{Crampin,Hayashi1}. 

Assuming the commutativity $w_{12}=w_{21}$, 
we have 
\begin{subequations} 
\begin{align}
w_{1, x}+w_{0, x}&=a_1 \sinh(w_{1}-w_{0}),
\label{4e3}\\
w_{2, x}+w_{0, x}&=a_2 \sinh(w_{2}-w_{0}),
\label{4e4}\\
w_{12, x}+w_{1, x}&=a_2 \sinh(w_{12}-w_{1}),
\label{4e5}\\  
w_{12, x}+w_{2, x}&=a_1 \sinh(w_{12}-w_{2}).
\label{4e6}   
\end{align} 
\end{subequations}

\vspace*{-4mm}
\noindent
Manipulating 
Eq.(\ref{4e3})$-$Eq.(\ref{4e4})$-$Eq.(\ref{4e5})$+$Eq.(\ref{4e6}), 
derivative terms are canceled out, 
so that we have an algebraic relation
\begin{equation}
\tanh\Big(\frac{w_{12}-w_{0}}{2}\Big)=-\frac{a_1+a_2}{a_1-a_2} 
   \tanh\Big(\frac{w_{1}-w_{2}}{2}\Big).  
\label{4e7}
\end{equation}
This equation satisfies Eqs.(\ref{4e3})-(\ref{4e6}), so that 
$w_{12}$ can be new solution. 
Notice that from the time-dependent
1-soliton solutions $w_0$, $w_1$, and $w_2$, we obtain the time-dependent new
solution $w_{12}$, so that Eq.(\ref{4e2}) is not necessary to construct 
the new solution.
By using the above B\"{a}cklund transformation, we can 
construct a new soliton solution $w_{12}$ from  
1-soliton solutions $w_1$,$w_2$, and $w_0$. 

Taking that $w_0=0$ is a trivial solution into account,  
we have 
2-soliton solutions $w_{12}$, and $w_{13}$ by using 
1-soliton solutions $w_1$, 
$w_2$, and $w_3$ through  
$\tanh w_i/2={\rm e}^{X_i}$ with 
$X_i=a_i x+a_i^3 t+c_i$, 
\begin{align}
\tanh\Big(\frac{w_{12}}{2}\Big)&=
-\frac{1}{a_{12}} \tanh\Big(\frac{w_{1}-w_{2}}{2}\Big)=
 -\frac{1}{a_{12}} \frac{\tanh(w_{1}/2)-\tanh(w_{2}/2)}
{1-\tanh(w_{1}/2) \tanh(w_{2}/2)},
\label{4e8}\\
\tanh\Big(\frac{w_{13}}{2}\Big)&=
-\frac{1}{a_{13}} \tanh\Big(\frac{w_{1}-w_{3}}{2}\Big)=
 -\frac{1}{a_{13}} \frac{\tanh(w_{1}/2)-\tanh(w_{3}/2)}
{1-\tanh(w_{1}/2) \tanh(w_{3}/2)},
\label{4e9}
\end{align}
with $a_{ij}=(a_i-a_j)/(a_i+a_j)=-a_{ji}$. 

Next, let us construct a 3-soliton solution. 
Assuming the commutativity 
$w_{123}=w_{132}$, we have
\begin{equation}
\tanh\Big(\frac{w_{123}-w_{1}}{2}\Big)=-\frac{a_2+a_3}{a_2-a_3} 
   \tanh\Big(\frac{w_{12}-w_{13}}{2}\Big).
\label{4e10}
\end{equation}
We express the above with $ t_1=\tanh(w_{1}/2)$,
$t_2=\tanh(w_{2}/2)$, and $t_3=\tanh(w_{3}/2)$, and 
\begin{equation}
\tanh\left(\frac{w_{123}}{2}\right)=\frac{g_{123}}{f_{123}}, \quad 
\mbox{with}\quad f_{123}
=\sum_{i=0}^{7} c_i p_i(t), 
\quad  g_{123}=\sum_{i=0}^{7} c_i q_i(t). 
\label{4e11}
\end{equation}
In the expression above, we denote 
$$
 c_0=a_{12} a_{13} a_{23},
\quad c_1=-a_{12}+a_{13}-a_{23},
\quad c_2=-a_{12} a_{13} a_{23} -a_{13}+a_{23},
$$
$$
c_3=-a_{12} a_{13} a_{23}+a_{12}+a_{23}, 
\quad
c_4=-a_{23},
\quad c_5=a_{12},
\quad  c_6=-a_{13},
\quad  
c_7=a_{12}a_{13} a_{23} -a_{12}+a_{13}, 
$$
$$
p_0=1,
\quad p_1=t_1^2,
\quad p_2=t_1 t_2,
\quad p_3=t_1 t_3,
\quad p_4=t_2 t_3,
\quad p_5=t_1^3 t_2,
\quad p_6=t_1^3 t_3,
\quad p_7=t_1^2 t_2 t_3 , 
$$
$$
q_0=t_1^3 t_2 t_3,
\quad q_1=t_1 t_2 t_3,
\quad q_2=t_1^2 t_3,
\quad q_3=t_1^2 t_2,
\quad q_4=t_1^3,
\quad q_5=t_3,
\quad q_6=t_2,
\quad q_7=1, 
$$
which satisfy $p_i q_i =t_1^3 t_2 t_3\ (i=0, 1, \cdots, 7)$. 
We can observe that $\tanh(w_{123}/2)$
 is not cyclic symmetric in $t_1$, $t_2$, and $t_3$. 
This is the non-cyclic symmetric 3-soliton solution of the 
mKdV equation derived from 
another mKdV type B\"{a}cklund transformation.

The non-cyclic symmetric 3-soliton solution for the sinh-Gordon
equation can be obtained by replacing
$\tanh(w_{123}/2) \rightarrow \tanh(\theta_{123}/4)$ and
$t_i=\tanh w_i/2={\rm e}^{.X_i} \rightarrow 
\hat{t}_i=\tanh \theta_i/4={\rm e}^{\hat{X}_i}$.
We can connect the mKdV equation with the 
sinh-Gordon equation in another way. If we put $w=c_1$ in 
Eq.(\ref{4e2}), we have $w'_x=a \sinh(w'-c_1)$ and
$w'_t=a^3 \sinh(w'-c_1)$, which gives the sinh-Gordon
equation $\Theta_{xt}=a^4 \sinh(\Theta)$ through 
the relation $\Theta=2(w'-c_1)$, and the $a$-dependence
can be eliminated by the redefinition of  $x \rightarrow x/a$, 
and $t\rightarrow t/a^3$.

%%%%%%%%%%%%
\section{Summary and discussions} 
\setcounter{equation}{0}

We consider the reason why special non-linear differential
 equations, such as  
 KdV/mKdV/sinh-Gordon equations, 
have the systematic $N$-soliton solution is because such soliton 
equations have SO(2,1) $\cong$ GL(2,$\mathbb R$) $\cong$ M\"{o}bius group 
structure.
The systematic $N$-soliton solutions are given as the result of the addition 
formula of these Lie groups. 
As the representation of the addition formula of the 
Lie groups, the algebraic function such as trigonometric/hyperbolic/elliptic
functions appear. 

We have studied to unify the soliton system through 
the common addition formula coming from the common 
KdV type Hirota form B\"{a}cklund transformation
$D_{x}^2 (f'\pm g')\cdot (f \pm g)=a^2 (f'\pm g')(f \pm g)/4$,   
which is equivalent to the well-known
KdV type B\"{a}cklund transformation 
$z'_x+z_x=-a^2/2+(z'-z)^2/2$ 
where $z=-2[\log(f\pm g)]_x$, $z'=-2[\log(f'\pm g')]_x$.
If we construct the $N$-soliton solutions 
through the KdV type B\"{a}cklund transformation, we can transform 
different KdV/mKdV/sinh-Gordon equations and B\"{a}cklund 
transformations of the standard form into the same common Hirota form 
 and B\"{a}cklund 
transformation,  Eq.(\ref{2e9}), Eq.(\ref{2add4}), Eq.(\ref{2e21}), Eq,(\ref{2e17}) 
and Eq.(\ref{2e39}) except the equation which has the time-derivative term. 
In KdV/mKdV equation, the equation which has the time-derivative 
term becomes the same Eq.(\ref{2add3}) and Eq.(\ref{2e16}) but it is different 
from sinh-Gordon's one Eq.(\ref{2e38}). The difference is only the 
time-dependence and the main 
structure of the $N$-soliton solutions
has the same common form for KdV/mKdV/sinh-Gordon systems. 
Then the $N$-soliton solutions for the sinh-Gordon equation 
is obtained just by the replacement $a_i x+a_i^3 t \rightarrow a_i x+t/a_i$
from KdV/mKdV $N$-soliton solutions.

We have also given the general addition formula of this common
KdV type Hirota form B\"{a}cklund transformation.
This addition formula is applicable also to construct the 
elliptic $N$-soliton solutions and there will be no other way 
to construct $N$-soliton solutions for the elliptic case~\cite{Hayashi2}.
Then it is useful to construct $N$-soliton solutions for 
trigonometric/hyperbolic/elliptic soliton solutions.

While by using another mKdV/sinh-Gordon type 
B\"{a}cklund transformation $w'_x+w_x=a \sinh (w'-w) $, we have
the non-cyclic symmetric solution. 
For the non-cyclic symmetric $N$-soliton solutions for the KdV equation, 
we can construct that from the mKdV non-cyclic symmetric $N$-soliton
solutions through the Miura transformation $u=\pm v_x+v^2$. 
We have given the explicit non-cyclic symmetric 3-soliton solutixon for KdV/mKdV/sinh-Gordon equations.
In the case of the mKdV type B\"{a}cklund, we 
add the comment to connect the mKdV equation with the sinh-Gordon equation
at the end of section 4.

We clarify what kind of Hirota type KdV/mKdV/sinh-Gordon equations correspond 
to the KdV type or the mKdV type B\"{a}cklund transformations.
Eq.(2.18a) and Eq.(2.18b) are equations for the KdV
type B\"{a}cklund transformation and Eq.(2.17) is the equation
for the mKdV type B\"{a}cklund transformation.

We expect that the higher rank Lie groups 
and higher genus algebraic functions appear in the higher dimensional 
and the higher symmetric soliton system.

\vspace{10mm}

%%%%%%%%%%%%%%%%%%%%%%%%%%%%%%%%%%%


\begin{thebibliography}{99}
\bibitem{Gardner}
C.S. Gardner, J.M. Greene, M.D. Kruskal, and R.M. Miura, 
Phys. Rev. Lett. {\bf 19}, 1095 (1967).
\bibitem{Lax} 
P.D. Lax, Commun. Pure and Appl. Math. {\bf 21}, 467 (1968).
\bibitem{Zakhrov}
V.E. Zakharov and A.B. Shabat, Sov. Phys. JETP {\bf 34}, (1972) 62.
\bibitem{Ablowitz}
M.J. Ablowitz, D.J. Kaup, A.C. Newell, and H. Segur, 
Phys. Rev. Lett. {\bf 31}, 125 (1973).  
\bibitem{Sasaki}
R. Sasaki, Nucl. Phys. {\bf B154}, 343 (1979).  
\bibitem{Wahlquist}
H.D. Wahlquist and F.B. Estabrook, Phys. Rev. Lett. {\bf 31}, 1386 (1973). 
\bibitem{Wadati1} 
M. Wadati, J. Phys. Soc. Jpn. {\bf 36}, 1498 (1974). 
\bibitem{Wadati2}
K. Konno and M. Wadati, Prog. Theor. Phys. {\bf 53}, 1652 (1975). 
\bibitem{Hirota1}
R. Hirota, Prog. Theor. Phys. {\bf 52}, 1498 (1974).  
\bibitem{Hirota2}
R. Hirota, Phys. Rev. Lett. {\bf 27}, 1192 (1971). 
\bibitem{Hirota3}
R. Hirota, J. Phys. Soc. Jpn. {\bf 33}, 1456 (1972). 
\bibitem{Hirota4}
R. Hirota, J. Phys. Soc. Jpn. {\bf 33}, 1459 (1972).  
\bibitem{Hirota5}
R. Hirota, {\it Direct Method in Soliton Theory} (Cambridge 
University Press, Cambridge, 2004). 
\bibitem{Sato}
M. Sato, RIMS Kokyuroku (Kyoto University) {\bf 439}, 30 (1981).
\bibitem{Lepowsky}
J. Lepowsky and R.L. Wilson, Comm. Math. Phys. {\bf 62}, 43 (1978).
\bibitem{Date}
E. Date, M. Kashiwara, and T. Miwa, Proc. Japan Acad. {\bf 57A}, 387 (1981).
\bibitem{Weiss}
J. Weiss, J. Math. Phys. {\bf 24}, 1405 (1983).
\bibitem{Wazwaz}
A-M. Wazwaz, {\it Partial Differential Equations and Solitary Waves
Theory},
 (Springer-Verlag, Berlin Heidelberg, 2009). 
\bibitem{Kaur1-1}
L. Kaur and A-M. Wazwaz,
Romamian Reports in Physics {\bf 71}, 102 (2019).
\bibitem{Kaur1-2}
L. Kaur and A-M. Wazwaz, Physica Scripta {\bf 93}, 075203 (2018).
\bibitem{Kaur1-3}
L. Kaur and A-M. Wazwaz, Nonlinear Dynamics {\bf 94}, 2469 (2018).
\bibitem{Kaur2}
L. Kaur and R.K. Gupta,  Mathematical Methods
in the Applied Sciences {\bf 36}, 584 (2013).
\bibitem{Hayashi1}
M. Hayashi, K. Shigemoto, and T. Tsukioka, 
Mod. Phys. Lett. {\bf A34}, 1950136 (2019).
\bibitem{Hayashi2}
M. Hayashi, K. Shigemoto, and T. Tsukioka, 
J. Phys. Commun. {\bf 3}, 045004 (2019).
\bibitem{Hayashi3}
M. Hayashi, K. Shigemoto, and T. Tsukioka, 
J. Phys. Commun.  {\bf 3}, 085015 (2019).
\bibitem{Bianchi}
L. Bianchi, {\it Vorlesungen \"{u}ber Differenzialgeometrie} 
(Teubner, 1899), p.418. 
\bibitem{Rogers1}
C. Rogers and W.F. Shadwick, {\it B\"{a}cklund Transformation and their Applications},
 (Academic Press, Inc, NewYork, 1982).
\bibitem{Rogers2}
C. Rogers and W.K. Schief, 
{\it B\"{a}cklund and Darboux Transformations:
Geometry and Modern Applications in Soliton Theory},
 (Cambridge University Press, Cambridge, 2002).%
\bibitem{Bowcock}
P. Bowcock, E. Corrigan and C. Zambon, Int. J. Mod. Phys. 
.{\bf A19(Supplement Issue 2)}, p.82-91 (2004).
\bibitem{Corrigan}
E. Corrigan and C. Zambon, J. Phys. .{\bf A42}, 475203 (2009).
\bibitem{Gomes}
J.F. Gomes, A.L. Retore and A.H. Zimerman, J. Phys.{\bf A49}, 504003.
(2016).
\bibitem{Spano}
N.I. Spano, A.L. Retore, J.F. Gomes, A.R. Aguirre and A.H. Zimerman,
 ``The sinh-Gordon defect matrix generalized for n defects'', in 
{\it Physical and Mathematical Aspects of Symmetries. 
Proceedings of the 31st International Colloquium in Group Theoretical 
Methods in Physics, p.73-78} (New York: Springer, 2017), 
{\tt [arXiv:1610.01856 [nlin.SI]]}.%
\bibitem{Shigemoto1}
K. Shigemoto, ``The Elliptic Function in Statistical Integrable Models'', 
Tezukayama Academic Review {\bf 17}, 15 (2011), 
{\tt [arXiv:1603.01079v2[nlin.SI]]}. 
\bibitem{Shigemoto2}
K. Shigemoto, 
``The Elliptic Function in Statistical Integrable Models II'', 
Tezukayama Academic Review {\bf 19}, 1 (2013), 
{\tt [arXiv:1302.6712v1[math-ph]]}. 
\bibitem{Crampin}
M. Crampin, Phys. Lett. {\bf A66}, 170 (1978).
\end{thebibliography}
\end{document}